\documentclass[conference]{IEEEtran}
\IEEEoverridecommandlockouts
\usepackage{cite}
\usepackage{amsmath,amssymb,amsfonts}
\usepackage{algorithmic}
\usepackage{graphicx}
\usepackage{textcomp}
\usepackage{xcolor}
\usepackage{soul}

\def\BibTeX{{\rm B\kern-.05em{\sc i\kern-.025em b}\kern-.08em
    T\kern-.1667em\lower.7ex\hbox{E}\kern-.125emX}}
\usepackage[rawfloats=true]{floatrow} 
\restylefloat{figure}

\begin{document}
\sloppy

\makeatletter
\newcommand{\linebreakand}{%
  \end{@IEEEauthorhalign}
  \hfill\mbox{}\par
  \mbox{}\hfill\begin{@IEEEauthorhalign}
}
\makeatother

\title{A Coding Theory Perspective on Multiplexed Molecular Profiling of Biological Tissues}

\author{
\IEEEauthorblockN{Luca D'Alessio\IEEEauthorrefmark{1}}
\IEEEauthorblockA{
\textit{Broad Institute, Cambridge, MA}\\
ldalessi@broadinstitute.org}
\and
\IEEEauthorblockN{Litian Liu\IEEEauthorrefmark{1}}
\IEEEauthorblockA{
\textit{MIT, Cambridge, MA}\\
litianl@mit.edu}
\and
\IEEEauthorblockN{Ken Duffy}
\IEEEauthorblockA{
\textit{Maynooth University, Ireland}\\
ken.duffy@nuim.ie}
\and
\IEEEauthorblockN{Yonina C. Eldar}
\IEEEauthorblockA{
\textit{Weizmann Institute of Science, Israel}\\
yonina.eldar@weizmann.ac.il}
\linebreakand
\IEEEauthorblockN{Muriel M\'edard}
\IEEEauthorblockA{
\textit{MIT, Cambridge, MA}\\
medard@mit.edu}
\and
\IEEEauthorblockN{Mehrtash Babadi}
\IEEEauthorblockA{
\textit{Broad Institute, Cambridge, MA}\\
mehrtash@broadinstitute.org}
\thanks{Luca D'Alessio and Mehrtash Babadi acknowledge funding and support from Data Sciences Platform (DSP), Broad Institute. Litian Liu acknowledges financial support from Klarman Family Foundation. Yonina Eldar acknowledges funding from NIMH grant 1RF1MH121289-0. The authors thank Samouil L. Farhi for beneficial discussions, and Aviv Regev for supporting this project.}}

\setlength{\textfloatsep}{5pt}

\maketitle
\begingroup\renewcommand\thefootnote{*}
\footnotetext{These two authors contribute equally to the work.}
\endgroup

\newcommand{\prob}{\mathrm{Pr}}

\begin{abstract}
High-throughput and quantitative experimental technologies are experiencing rapid advances in the biological sciences. One important recent technique is multiplexed fluorescence {\em in situ} hybridization (mFISH), which enables the identification and localization of large numbers of individual strands of RNA within single cells. Core to that technology is a coding problem: with each RNA sequence of interest being a codeword, how to design a codebook of probes, and how to decode the resulting noisy measurements? Published work has relied on assumptions of uniformly distributed codewords and binary symmetric channels for decoding and to a lesser degree for code construction. Here we establish that both of these assumptions are inappropriate in the context of mFISH experiments and substantial decoding performance gains can be obtained by using more appropriate, less classical, assumptions. We propose a more appropriate asymmetric channel model that can be readily parameterized from data and use it to develop a {\em maximum a posteriori} (MAP) decoders. We show that false discovery rate for rare RNAs, which is the key experimental metric, is vastly improved with MAP decoders even when employed with the existing sub-optimal codebook. Using an evolutionary optimization methodology, we further show that by permuting the codebook to better align with the prior, which is an experimentally straightforward procedure, significant further improvements are possible.
\end{abstract}

\section{Introduction}\label{sec:intro}

In recent years, the field of single-cell biology has witnessed transformative advances in experimental and computational methods. Of particular interest is the recent advent of multiplexed fluorescence {\em in situ} (in-place) hybridization (mFISH) microscopy techniques that allow molecular profiling of hundreds of thousands of cells without disturbing their complex arrangement in space. This highly-informative data modality paves the way to transformative progress in many areas of biology, including understanding morphogenesis, tissue regeneration, and disease at molecular resolution.

One of the major challenges in designing such experiments is the vastness of functional bio-molecules. For example, the human genome codes nearly 30k non-redundant types of RNA molecules, many of which translate to proteins with specific functions. Modern data-driven biology heavily relies on our ability to measure as many different types of functional molecules as possible. Clearly, a sequential imaging approach is impractical. Fortunately, a typical cell produces a rather sparse set of all molecules, and some of the most promising mFISH techniques exploit molecular sparsity in space together with coding ideas in order to multiplex the measurements into fewer imaging rounds~\cite{chen2015spatially,moffitt2016highthroughput,moffitt2016highperformance,shah2017seqfish,eng2019transcriptome}. 

In brief, the mFISH technique involves assigning binary codes to RNA molecules of interest, chemically synthesizing and ``hybridizing'' these codes to the molecules, and measuring them in space one bit at a time via sequential fluorescence microscopy. A more detailed account of one such pioneering technique known as MERFISH (``multiplexed error-robust fluorescence {\em in situ} hybridization'')\cite{chen2015spatially} is given in Sec.~\ref{sec:merfish} (also, cf. Fig.~\ref{fig:merfish_overview}). An important part of the MERFISH protocol is the utilization of sparse codes with large minimum distance to allow error correction. Referred to as MHD4 codes~\cite{chen2015spatially,moffitt2016highthroughput,moffitt2016highperformance}, these 16-bit codes have minimum Hamming distance 4 and contain 4 ones and 12 zeros each. The bit imbalance is motivated by the empirically observed $\sim 2\times$ higher signal fallout rate compared to false alarm. There are only 140 such codes and therefore one is limited to measuring at most 140 distinct molecules. These codes are randomly assigned to the RNA molecules of interest. The decoding method in current use relies on quantization, Hamming error correction, and rejection of ambiguous sequences.

We point out that the assumptions motivating the codebook construction and decoding, tacitly yet heavily, rely on source uniformity and to a certain extent on the binary {\em symmetric} channel paradigm, both of which are violated in the context molecular profiling. For channel coding in communication, source can be readily assumed as uniformly distributed thanks to compression in source coding and the separation theorem \cite{shannon1948mathematical}. In molecular profiling, however, source compression is not applicable and the distribution of RNA molecules is {\em extremely} non-uniform. Moreover, fluorescence microscopy is established to be highly asymmetric in terms of fallout and false alarm. These violated assumptions become a source of potential problems when directly applying communication encoding and decoding paradigms. For example, the false discovery rate of rare molecules is found to be unacceptably high in replicate experiments~\cite{chen2015spatially,moffitt2016highthroughput}, which we later show to be a consequence of the assumed source uniformity. Accurate quantification of rare RNA molecules (e.g. transcription factors) is particularly important for data-driven biological discovery since rare molecules often signal rare events, transient cells states, etc. This motivates our primary goal in this paper: to incorporate the prior non-uniformity in the decoding process in a principled way in order to control false discovery rate of rare molecules. In practice, either accurate priors are known, can be estimated from the data, or can be measured cheaply and effortlessly (e.g. using bulk RNA sequencing \cite{chen2015spatially}).

This paper is organized as follows: we review the MERFISH protocol in Sec.~\ref{sec:merfish} and propose a generative model for the data in Sec.~\ref{sec:generative}, along with a model fitting algorithm and a procedure to derive a more tractable binary asymmetric channel (BAC) formulation from the fitted model. The BAC framework allows us to evaluate the performance of different encoding and decoding schemes. We incorporate the prior non-uniformity into the decoding algorithm by developing a {\em maximum a posteriori} (MAP) decoder with a tunable rejection threshold in Sec.~\ref{sec:decoding}. We show that the false discovery rate of rare RNAs, which is the key experimental metric, is vastly improved compared to the presently used MLE-based decoding method~\cite{chen2015spatially,moffitt2016highthroughput,moffitt2016highperformance}, even when employed with the existing sub-optimal MHD4 codebook. Finally, we take a first step in data-driven code construction in Sec.~\ref{sec:encoding}. Using an evolutionary optimization methodology, we show that by permuting the codebook to better align with the prior, which is an experimentally straightforward procedure, significant further improvements are possible. We conclude the paper in Sec.~\ref{sec:conclusion} with of follow up research directions.

\subsection{A brief overview of the MERFISH protocol}\label{sec:merfish}

\begin{figure}
    \centering
    \includegraphics[scale=0.70]{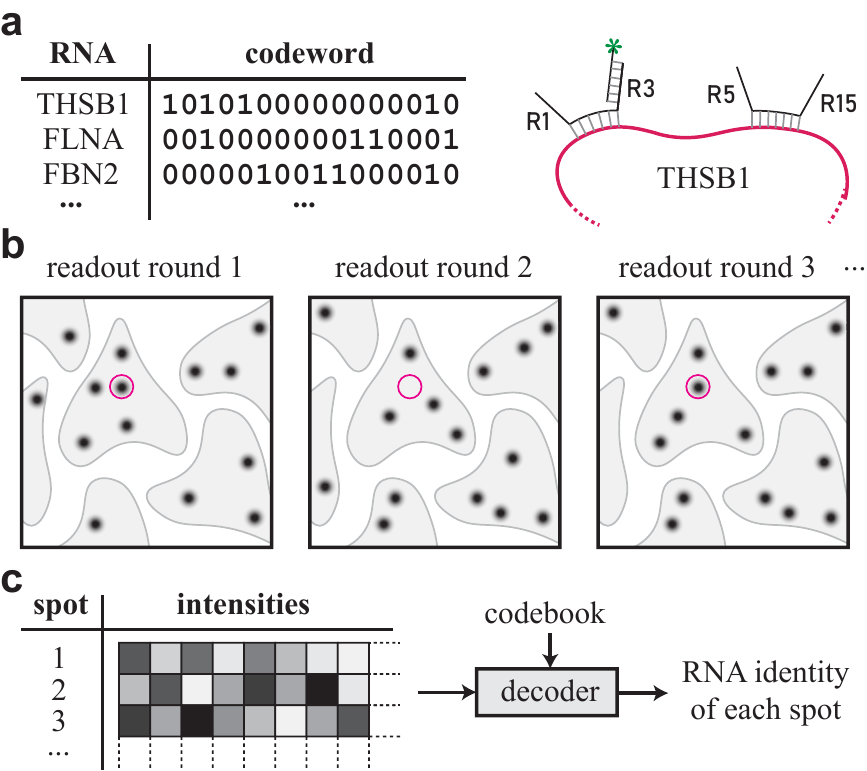}
    \caption{A schematic overview of a typical mFISH experiment. (a) codebook and probe design; (b) sequential imaging; (c) image processing and decoding.}
    \label{fig:merfish_overview}
\end{figure}

In this section, we briefly review the MERFISH protocol~\cite{chen2015spatially}, recount different sources of noise and nuisance, and motivate a generative process for MERFISH data. Fig.~\ref{fig:merfish_overview} shows a schematic overview of the MERFISH technique. This protocol consists of four main steps: {\bf Step~1.} A unique binary codeword of length $L=16$ is assigned to the RNA molecules of interest; {\bf Step~2.} The specimen is stained with carefully designed short RNA sequences called {\em encoding probes}. The middle part of the encoding probes bind with high specificity to a single RNA type while their flanking heads and tails contain a subset of $L$ artificial sequences, $\{R_1, \ldots, R_L\}$, called {\em readout sequences}. The choice of readout sequences reflects the intended binary codeword. For instance, if the code for a certain RNA type contains ``1'' at positions 1, 3, 5, and 15, the encoding probes are designed to have $R_1, R_3, R_5$, and $R_{15}$ flanking sequences (see Fig.~\ref{fig:merfish_overview}a); {\bf Step~3.} The prepared tissue undergoes $L$ rounds of imaging. Imaging round $l$ begins with attaching {\em fluorescent readout probes for round {l}} to the prepared tissue. These probes bind to the flanking readout sequences and contain a fluorescent dye that emits light upon excitation. The round ends with bleaching the dye. In effect, imaging round $l$ reveals the position of all RNA molecules having ``1'' in their binary codeword at position $l$. {\bf Step~4.} Finally, the position of RNA molecules, which appear as bright spots, are identified using conventional image processing operations (see Fig.~\ref{fig:training_data}). The data is summarized as an $N \times L$ intensity matrix ($N$ being the number of identified spots) and is ultimately {\em decoded} according to the codebook. MERFISH measurements are affected by several independent sources of noise. These include factors that are intrinsic to individual molecules, such as (1) stochasticity in the hybridization of encoding and readout probes, (2) random walk of molecules between imaging rounds, and (3) CCD camera shot noise. These factors module the intensity measurements independently in each round and are largely uncorrelated across rounds. Extreme multiplexing (e.g. as in the seqFISH+ protocol~\cite{eng2019transcriptome}) further leads to {\em interference noise} due to signal mixing between nearby molecules. This nuisance, however, is rather negligible in the MERFISH protocol.

\section{Methodology and Results}\label{sec:method}
\vspace{1.6em}

\subsection{A generative model for mFISH data}\label{sec:generative}
\vspace{1.6em}

\begin{figure}
    \centering
    \includegraphics[scale=0.43]{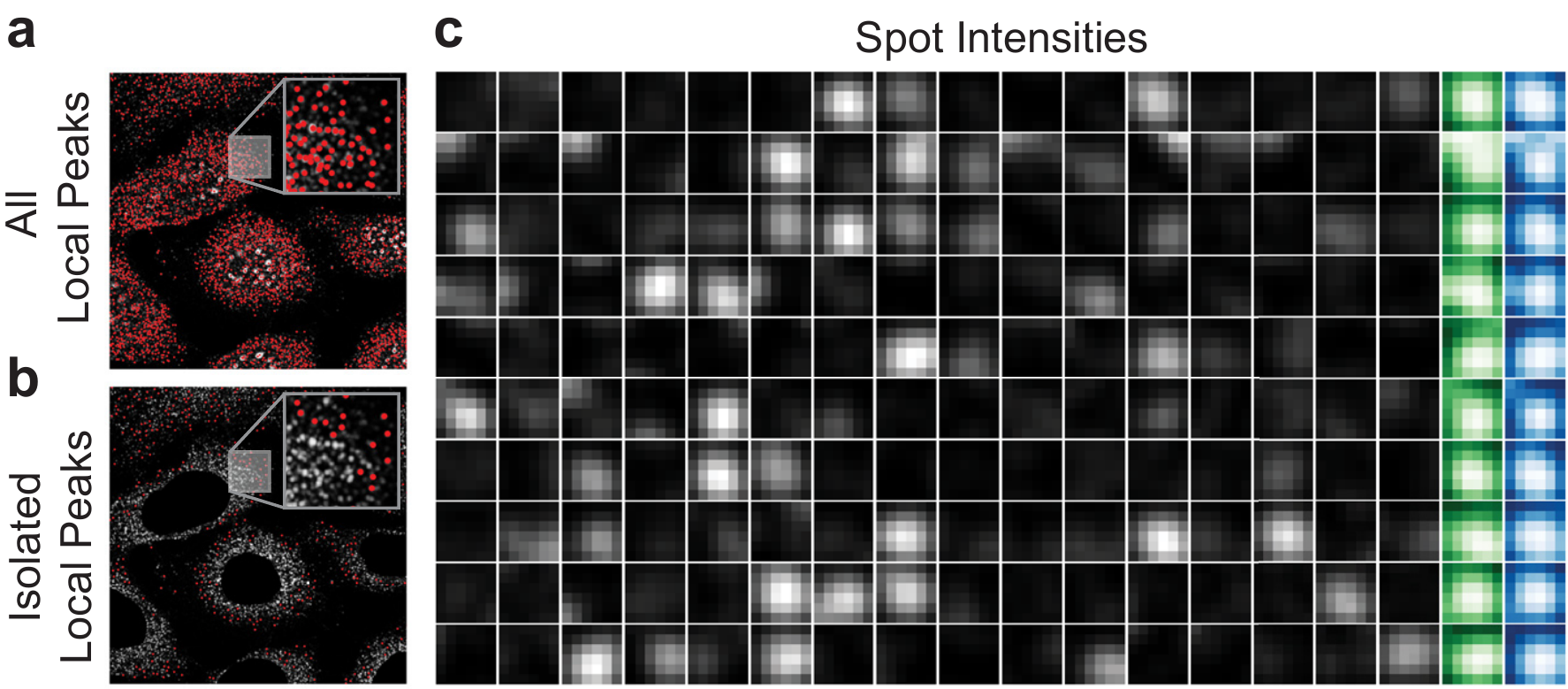}
    \caption{Extraction of isolated spots from MERFISH images (data from \cite{moffitt2016highthroughput}). (a) local peak finding; (b) identification of isolated spots; (c) intensity series from 10 random spots (rows); the leftmost 16 columns show the intensity measurements; the last two column show the summed intensity and nearest-neighbor cross-correlations and are used for filtering of poorly localized spots.}
    \label{fig:training_data}
\end{figure}

\begin{figure}
    \centering
    \includegraphics[scale=0.50]{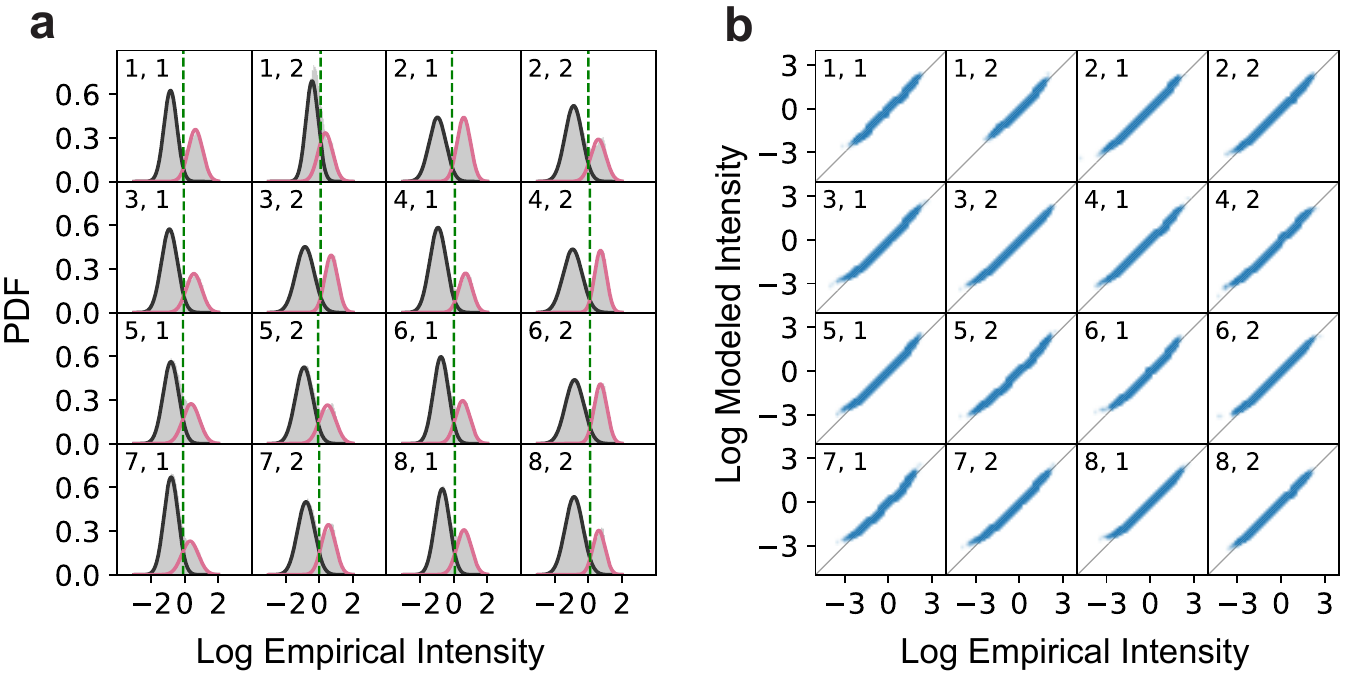}
    \caption{Modeling spot intensities as two-component Gaussian mixture for each data dimension (i.e. readout round and color channel). (a) model fitting (black and red lines) and empirical histograms (gray); the green lines indicate the quantization thresholds for the ensuing BAC approximation; (b) QQ-plots for each data dimension; the labels shown in the sub-panels indicate hybridization rounds $\{1, \ldots, 8\}$ and color channels $\{1, 2\}$.}
    \label{fig:gmm_fit}
\end{figure}

In this section, we present a simple generative model for MERFISH spot intensity data, fit the model to real data, and evaluate the goodness of fit. This model will serve as a foundation for developing a MAP decoder. Fig.~\ref{fig:training_data} shows a typical example of MERFISH data from~\cite{moffitt2016highthroughput}. We formalize the data generating process as follows: let $\mathsf{C} \subset \{0, 1\}^L$ be a set of codewords with cardinality $|\mathsf{C}| = K$ which are assigned to $G \leq K$ molecules, let $a: \tilde{\mathsf{C}} \rightarrow \{1, \ldots, G\}$ be the bijective code assignment map where $\tilde{\mathsf{C}} \subset \mathsf{C}, |\tilde{\mathsf{C}}| = G$ is the set of used codes, and let $\boldsymbol{\pi}_{1:G}$ be the prior distribution of molecules. Setting aside interference effects, we model the fluorescence intensity series $\mathbf{I}_{1:L} \in [0, \infty)^{L}$ measured for an arbitrary molecule as follows:
\begin{equation}\label{eq:gen}
\begin{split}
    g &\sim \mathrm{Categorical}(\boldsymbol{\pi}),\\
    \mathbf{c} &= a^{-1}(g),\\
    \log I_l \, | \, c_l &\sim \mathcal{N}(\mu_l[c_l], \sigma^2_l[c_l]). 
\end{split}
\end{equation}
As discussed earlier, the intrinsic spot intensity noise (1) is multiplicative, (2) results from a multitude of independent sources, and (3) is uncorrelated across imaging rounds, motivating factorizing $\mathbf{I}_{1:L} \, | \, \mathbf{c}_{1:L}$ in $l$ and modeling each conditional as a Gaussian in the logarithmic space. The well-known heteroscedasticity of fluorescence noise is reflected in having two different $\sigma^2[c]$ for $c \in \{0, 1\}$ for the two binary states.

\subsection{Image processing and model fitting}
The most straightforward way to fit the generative model to empirical data is by observing that marginalizing the (discrete) molecule identity variable $g$ yields a two-component Gaussian mixture model (GMM) for $\log I_l$, with weights determined by the prior $\boldsymbol{\pi}$, codebook $\mathsf{C}$, and the assignment $a$. The model parameters $\{\mu_{1:L}[0], \sigma^2_{1:L}[0], \mu_{1:L}[1], \sigma^2_{1:L}[1]\}$ can be readily estimated by ML GMM fitting to each column of the spot intensity table (cf. Fig.~\ref{fig:merfish_overview}c), which can be performed efficiently using the conventional EM algorithm. In order to decouple the intrinsic and extrinsic spot noise in the raw data, we {\em censor} the dataset to only spatially isolated molecules. In brief, we process the images as described in \cite{chen2015spatially}, subtract the background, identify the position of molecules by local peak-finding, censor dense regions (e.g. cell nuclei), and retain local peaks that are separated from one another at least by $\sim 5~\mathrm{px}$, which is a few multiples of the diffraction limit. We perform additional filtering based on the spot intensity pattern and nearest-neighbor Pearson correlation (cf. Fig.~\ref{fig:training_data}) and only retain peaks with a symmetric appearance. This procedure yields $\sim$ 250k spots in the dataset published in \cite{moffitt2016highthroughput}. The obtained fits are shown in Fig.~\ref{fig:gmm_fit} along with QQ-plots that confirm a remarkably good fit to the empirical marginal histograms.

\subsection{Quantization, channel model and estimation}\label{label:bac}
The generative model specified by Eq.~\eqref{eq:gen} readily yields the posterior distribution $\prob(g\,|\,\mathbf{I};\boldsymbol{\pi}, \mathsf{C}, a)$ and can form the basis of an intensity-based MAP decoder. To make the formulation more amenable for computational and theoretical investigation, as well as making a connection to the currently used decoding method, we derive an approximate binary asymmetric channel (BAC) model from Eq.~\eqref{eq:gen} through quantization. The optimal quantization thresholds $\boldsymbol{\theta}_{1:L}$ are determined for each $l$ to be the point of equal responsibility between the two Gaussian components, i.e. $\sum_{g=1}^G \pi_g\, a^{-1}(g)[l]\,\mathcal{N}(\theta_{l} \, | \, \mu_l[1], \sigma_l^2[1]) = \sum_{g=1}^G \pi_g\, [1 - a^{-1}(g)[l]\,\mathcal{N}(\theta_{l} \, | \, \mu_l[0], \sigma_l^2[0])$, which amdits a closed-form solution. Here, $a$ and $\boldsymbol{\pi}$ correspond to the known code assignment and prior distribution of the data used for fitting. The fallout $p^{1 \rightarrow 0}$ and false alarm $p^{0 \rightarrow 1}$ rates are given by the integrated probability weights of the two Gaussian components below and above the threshold (cf. Fig.~\ref{fig:gmm_fit}{\bf a}), i.e. $p^{0 \rightarrow 1}_l = \Phi[(\mu_l[0] - \theta_l)/\sigma_l[0]]$ and $p^{1 \rightarrow 0}_l = \Phi[(\theta_l - \mu_l[1])/\sigma_l[1]]$, where $\Phi(\cdot)$ is the CDF of the standard normal distribution. We find $p_l^{0 \rightarrow 1}$ and $p_l^{1 \rightarrow 0}$ to be $0.046$ and $0.102$ (mean in $l$), respectively, for the data given in Ref.~\cite{moffitt2016highthroughput}, which is in agreement with the estimates reported therein. We, however, observed significant round-to-round variation in the channel parameters and as such, refrained from further simplifying the channel model to a single BAC for all imaging rounds $l$. We refer to the bundle of estimated BAC parameters as $\boldsymbol{\theta}_\mathrm{BAC}$.

\subsection{Decoding: MAP and MLE decoders}\label{sec:decoding}
A gratifying property of the BAC approximation of Eq.~\eqref{eq:gen} is allowing us to evaluate the performance of various decoding strategies without resorting to time-consuming simulations or further analytical approximations. In the BAC model, the likelihood of a binary sequence $\mathbf{x}_{1:L} \in \{0, 1\}^L$ conditioned on the codeword $\mathbf{c} \in \mathsf{C}$ is given as:
\begin{equation}
    \log \prob(\mathbf{x} \, | \, \mathbf{c}, \boldsymbol{\theta}_\mathrm{BAC}) = \sum_{l=1}^L \sum_{i,j \in \{0, 1\}}\, \delta_{c_l,i} \, \delta_{x_l, j} \, \log\, p_l^{i\rightarrow j},
\end{equation}
where $\delta_{\cdot,\cdot}$ is the Kronecker's delta function. We define the {\em posterior Voronoi set} for each codeword $\mathbf{c} \in \mathsf{C}$ as:
\begin{multline}
    \mathsf{V}(\mathbf{c} \, | \, a, \boldsymbol{\omega}, \mathsf{C}, \boldsymbol{\theta}_\mathrm{BAC}) = \big\{\mathbf{x} \in \{0, 1\}^L\, | \, \forall \mathbf{c}' \in \mathsf{C}, \mathbf{c} \neq \mathbf{c}':\\
    \omega_{a(\mathbf{c})} \, \prob(\mathbf{x} \, | \, \mathbf{c}, \boldsymbol{\theta}_\mathrm{BAC}) > \omega_{a(\mathbf{c}')} \, \prob(\mathbf{x} \, | \, \mathbf{c}', \boldsymbol{\theta}_\mathrm{BAC})\big\},
\end{multline}
where $\boldsymbol{\omega}_{1:G}$ is the {\em prior distribution assumed by the decoder}. The Voronoi sets are mutually exclusive by construction, can be obtained quickly for short codes by exhaustive enumeration, and determine the optimal codeword for an observed binary sequence. The {\em MLE decoder} corresponds to using a uniform prior, i.e. $\boldsymbol{\omega} \leftarrow \mathbf{1} / G$ whereas the {\em MAP decoder} corresponds to using the actual (non-uniform) prior governing the data $\boldsymbol{\omega} \leftarrow \boldsymbol{\pi}$. 
We additionally introduce a {\em MAP$_\mathrm{q}$ decoder}, which is a MAP decoder obtained from depleting the Voronoi sets from binary sequences for which the posterior probability of the best candidate code is below a set threshold $q$. Intuitively, the MAP$_\mathrm{q}$ decoder is a Bayesian decoder with {\em reject option} that trades precision gain for sensitivity loss by filtering dubious sequences from the Voronoi sets. The decoding algorithm introduced by Ref.\cite{chen2015spatially,moffitt2016highthroughput,moffitt2016highperformance} can be thought of as the MLE decoder with a rejection subspace given by $S_\mathrm{rej} = \{\mathbf{x} \, | \, \exists \, \mathbf{c}, \mathbf{c'}, \mathbf{c} \neq \mathbf{c}' \in \mathsf{C}: d_\mathrm{H}(\mathbf{c},\mathbf{x})=d_\mathrm{H}(\mathbf{c}',\mathbf{x})=d^*(\mathbf{x}, \mathsf{C})\}$ where $d_\mathrm{H}(\cdot,\cdot)$ is the Hamming distance and $d^*(\mathbf{x},\mathsf{C}) = \inf_{\mathbf{c} \in \mathsf{C}} d_\mathrm{H}(\mathbf{c}, \mathbf{x})$. We refer to this decoder as {\em Moffitt (2016)}. We remark that the acceptance criterion of Moffitt (2016) is extremely stringent: for MHD4 codes, $|S_\mathrm{acc}| = 9100$, which is only $\sim 13\%$ of all possible sequences (here, $S_\mathrm{acc}$ is the complement of $S_\mathrm{rej}$). In all cases, the {\em confusion matrix} $\mathcal{T}(\mathbf{c}\, | \, \mathbf{c}')$, i.e. the probability that a molecule coded with $\mathbf{c}'$ is decoded to $\mathbf{c}$, can be immediately calculated:

\begin{equation}
\mathcal{T}(\mathbf{c}\,|\,\mathbf{c}'; \boldsymbol{\pi}, \boldsymbol{\omega},  \boldsymbol{\theta}_\mathrm{BAC}) = \sum_{\mathbf{x} \in \mathsf{V}(\mathbf{c}|\boldsymbol{\omega}, \ldots)}\,\pi_{a(\mathbf{c}')}\,\prob(\mathbf{x} \, | \, \mathbf{c}',
\boldsymbol{\theta}_\mathrm{BAC})
\end{equation}
from which the marginal true positive rates $\mathrm{TPR}_{1:G}$ and false discovery rates $\mathrm{FDR}_{1:G}$ can be readily calculated.

\subsection{Comparing the performance of different decoders}

\begin{figure*}
    \centering
    \includegraphics[scale=0.38]{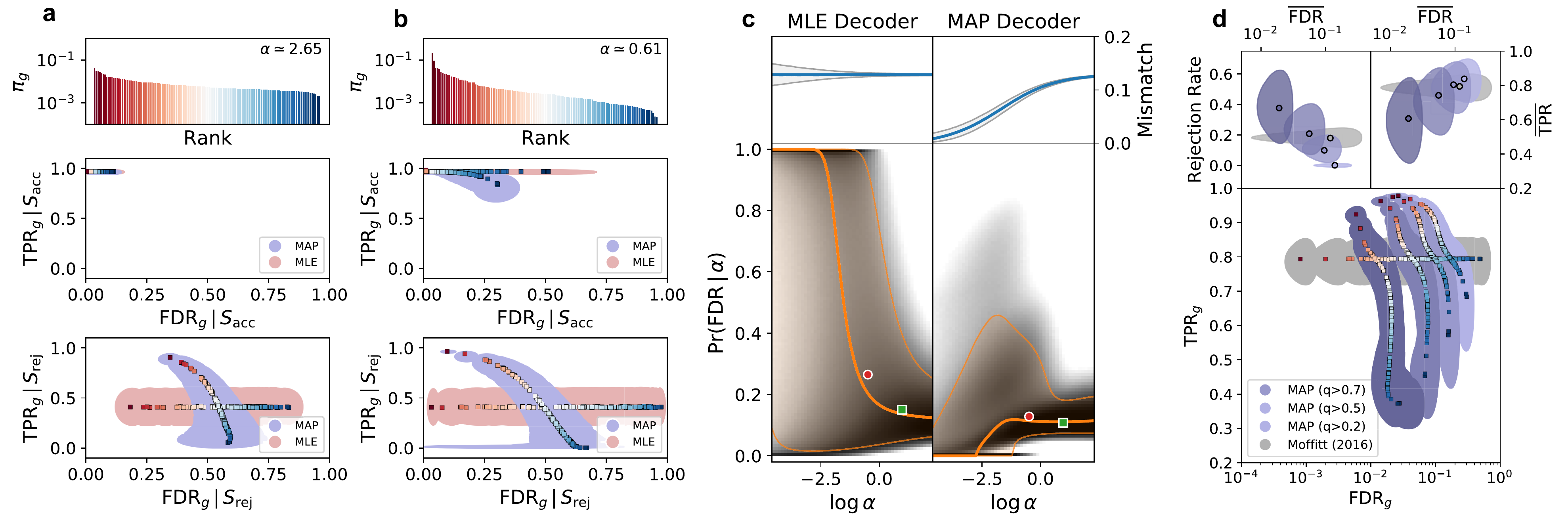}
    \vspace{0.4cm}
    \caption{Comparing the performance of different decoding schemes for randomly assigned MHD4 codes. (a) and (b) correspond to prior distribution for RNA molecules selected in \cite{moffitt2016highthroughput} and~\cite{moffitt2016highperformance}, respectively. The top panels show the rank-ordered prior distribution and the estimated Dirichlet concentration parameter $\alpha$; the middle and bottom panels show the marginal TPR and FDR for each molecule type conditioned on $S_\mathrm{acc}$ and $S_\mathrm{rej}$ subspaces (cf. Sec.~\ref{sec:decoding}); markers are color-coded according to prior rank of their corresponding molecules. Shaded regions indicate 5-95 percentile range as a matter of random code assignment; (c) the effect of prior non-uniformity on the performance of MLE and MAP decoders for randomly assigned MHD4 codes. The top panels show the uniform mismatch rate. The bottom panels show the histogram of marginal FDRs vs. Dirichlet prior concentration $\alpha$ in grayscale. The orange lines and regions indicate the median and 5-95 percentile ranges; (d) performance of MAP decoders with reject at different acceptance thresholds compared to method in \cite{moffitt2016highthroughput}.}
    \label{fig:emp_fdr_tpr}
\vspace{-1.6em}
\end{figure*}
Developments in previous sections allow us to compare the performance of MLE, MAP, MAP${}_q$, and Moffitt (2016). We use the BAC parameters obtained from the data in~\cite{moffitt2016highthroughput}, 16-bit MHD4 codes with random assignment, and two different previously estimated and published source priors with different degree of non-uniformity. As a first step, we compare the performance of our proposed MAP and MLE decoders separately inside $S_\mathrm{acc}$ and $S_\mathrm{rej}$, the acceptance and rejection subspaces of Moffitt (2016), in Fig.~\ref{fig:emp_fdr_tpr}{\bf a, b} (middle, bottom). The priors are shown on the top, including the estimated Dirichlet concentration $\alpha$. Marginal performance metrics for different molecules are color-coded according to their prior rank from red (most abundant) to blue (least abundant). The MLE decoder inside $S_\mathrm{acc}$ is equivalent to Moffitt (2016). Both decoders perform well in this subspace. While the MLE decoder is performing poorly inside $S_\mathrm{rej}$, providing a sound basis for rejection as in Moffitt (2016), the MAP decoder yields acceptable FDR, hinting that the $S_\mathrm{acc}$ is too stringent for the MAP decoder and better performance can be expected from MAP${}_q$. It also noticed that MAP decoder controls FDR much better than MLE inside $S_\mathrm{acc}$ for the more non-uniform prior. We explore this observation more systematically in panel {\bf c}. We sample $\boldsymbol{\pi}$ from a symmetric Dirichlet distribution with concentration $\alpha$ and calculate the distribution of the marginal FDRs (bottom) as well as the uniform mismatch rate (top). We notice that as the prior gets more concentrated $\log \alpha \rightarrow -\infty$, the MAP decoder behaves progressively better whereas the MLE decoder degrades and exhibits a bi-modal behavior: extremely low (high) FDR on abundant (rare) codes. As the prior gets more uniform $\log \alpha \rightarrow +\infty$, MLE and MAP become indistinguishable. The green and red symbols show the biological priors used in panels {\bf a} and {\bf b}, respectively, together with their estimated $\alpha$, in agreement with the trend of the Dirichlet prior model. Finally, panel {\bf d} compares the performance of the MAP${}_{q}$ decoder at different rejection thresholds $q$ with Moffitt (2016). The prior used here is the same as in panel {\bf b}. It is noticed that the MAP${}_q$ decoder is remarkably effective at controlling FDR for all codes whereas Moffitt (2016) degrades in FDR for rare codes, as expected from the source uniformity assumption. This finding explains the reportedly lower correlation between rare molecules in replicate experiments~\cite{chen2015spatially,moffitt2016highthroughput}. The smaller panels at the top of panel {\bf c} show mean TPR, FDR, and rejection rate across all molecules. The MAP${}_{0.5}$ decoder has similar sensitivity to Moffitt (2016) while yielding $\sim 20\%$ lower FDR on average and remarkably $\sim 60\%$ lower 5-95 FDR percentile range, implying significant improvement in reducing the mean and variance of false positives for both abundant and rare molecules.

\section{Data-driven code construction}\label{sec:encoding}
The results presented so far were obtained randomly assigning a fixed set of MHD4 codes. Constructing codes to better reflect channel asymmetry and prior non-uniformity is another attractive opportunity for improving the performance of mFISH protocols. Constructing application-specific codes for mFISH is outside the scope of the present paper and is a topic for future research. Here, we continue to thread on the theme of utilizing prior non-uniformity and show that optimizing the {\em assignment} of the even sub-optimal codes to molecules with respect to prior abundance can significantly reduce FDR. This is to be expected given the rather wide performance outcomes shown in Fig.~\ref{fig:emp_fdr_tpr} that result from random code assignment. Explicitly, we seek to optimize the scalarized metric $\overline{\mathrm{FDR}}(a, \boldsymbol{\pi}) = G^{-1} \sum_{g=1}^G \mathrm{FDR}_g (a, \boldsymbol{\pi})$ over the assignment operator $a$ for a given prior $\boldsymbol{\pi}$ through an evolutionary optimization process. We start with a population of $N=5000$ random code assignments, mutate the population via pairwise permutations with a small probability of $0.05$ per molecule per assignment, and select the fittest $N$ offsprings using $\overline{\mathrm{FDR}}$ as the measure of fitness. We do not use a crossover operation here. We hypothesize that a relevant surrogate for the optimality of $\overline{\mathrm{FDR}}$ is the concordance between the Hamming distance $d_H$ and the {\em prior distance} $d_{\boldsymbol{\pi}}(\mathbf{c}, \mathbf{c}') \equiv |\pi_{a(\mathbf{c})} - \pi_{a(\mathbf{c}')}|$. We investigate the emergence of this order by monitoring the following order parameter during the evolution:
\begin{equation}\label{eq:op}
\chi(a,\boldsymbol{\pi}) \equiv \frac{1}{G} \sum_{g=1}^G \rho^\mathrm{s}\Big[d_\mathrm{H}\big(a^{-1}(g), \mathbf{C}_a\big), d_{\boldsymbol{\pi}}\big(a^{-1}(g), \mathbf{C}_a\big)\Big],
\end{equation}
where $\rho^s[\cdot,\cdot]$ denotes the Spearman correlation and $\mathbf{C}_a$ is the ordered list of all codes used by $a$ over which the correlation is calculated. We refer to the population average of $\chi(a,\boldsymbol{\pi})$ as $\overline{\chi}$. We implement the evolutionary algorithm using the PyMOO package~\cite{pymoo} and vectorize the calculation of Voronoi sets with GPU acceleration. Fig.~\ref{fig:ga_optim} shows the results obtained by running the evolutionary optimization for three days (NVIDIA Testla P100 GPU, MHD4 codes, prior from \cite{moffitt2016highperformance}). Panel~{\bf a} shows the monotonic decline of $\overline{\mathrm{FDR}}$ to $\sim 75\%$ of its initial value (random assignment). This trend proceeds concurrently with a monotonic upturn in $\overline{\chi}$, providing evidence for the hypothesized matching order between $d_\mathrm{H}$ and $d_{\boldsymbol{\pi}}$. Panel~{\bf b} compares the performance metrics of the MAP decoder between the first and last population of code assignments. It is noticed that the optimized code assignment predominantly reduces $\overline{\mathrm{FDR}}$ of rare molecules, the mean FDR of which reduce to $\sim 50\%$ of randomly assigned codes. The possibility to reduce the FDR of rare molecules is a particularly favorable outcome in practice.

\begin{figure}
    \centering
    \includegraphics[scale=0.40]{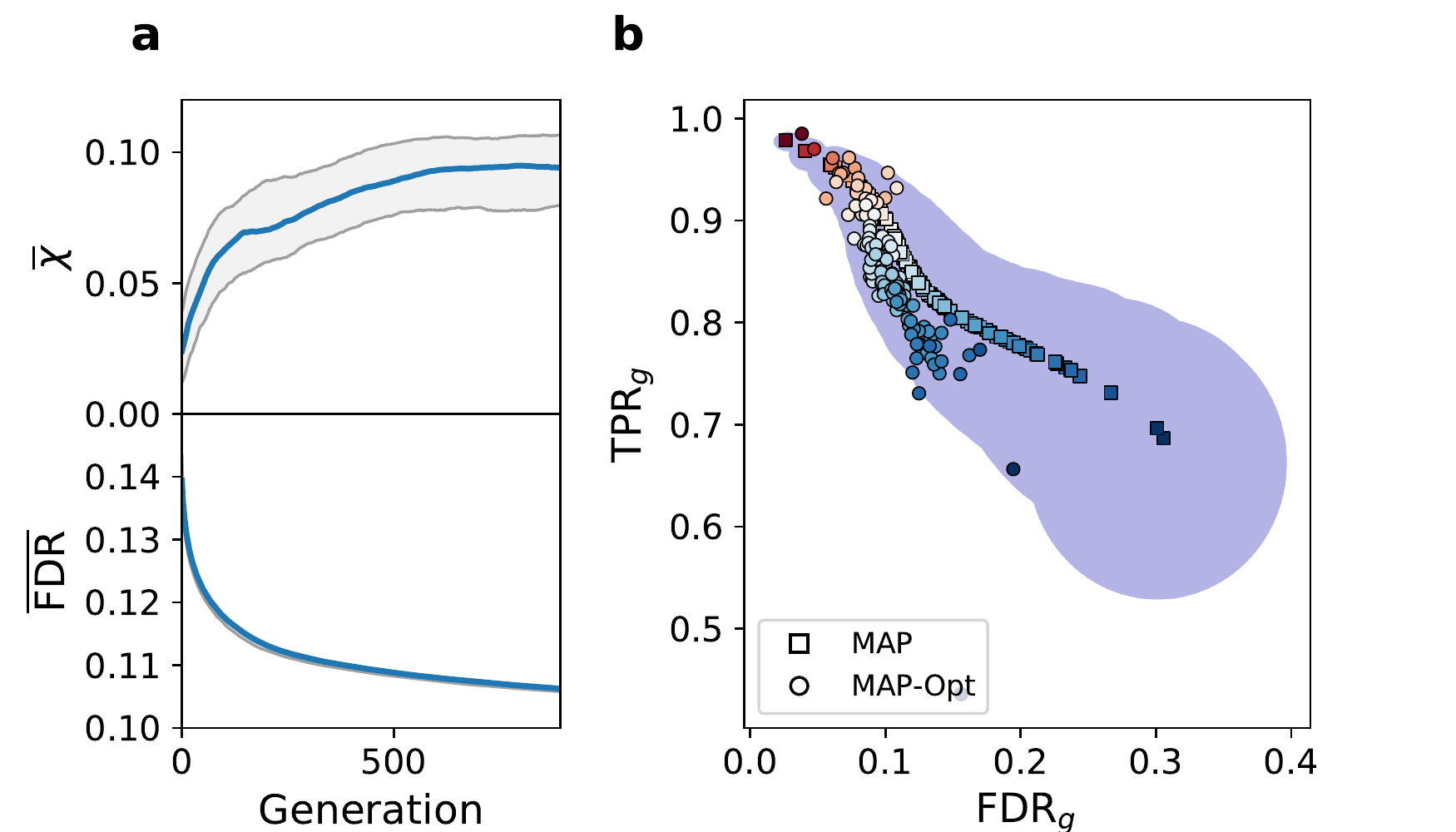}
    \caption{Evolutionary optimization of code assignment for MHD4 codes (for channel model described in Fig.~\ref{fig:gmm_fit} and prior distribution from \cite{moffitt2016highperformance}). (a) bottom: mean FDR vs. generation; top: $d_\mathrm{H}-d_{\boldsymbol{\pi}}$ matching order parameter vs. generation (see Eq.~\ref{eq:op}); (b) the performance of MAP decoder for randomly assigned codes (squares) vs. optimized assignment (circles).}
    \label{fig:ga_optim}
\end{figure}

\section{Conclusion and Outlook}\label{sec:conclusion}

In this paper, we reviewed multiplexed molecular profiling experiments from the perspective of coding theory, proposed a motivated generative model for the data, based on which we derived an approximate parallel BAC model for the system. We show that the exact MAP decoder of the BAC model vastly outperforms the decoding algorithm in current use in terms of controlling FDR of rare molecules, the key experimental metric. This is achieved by taking into account the non-uniformity of source prior, a ``non-classical'' aspect of multiplexed molecular profiling viewed as a noisy channel. We also took the first step in data-driven code construction and show that optimizing the assignment of existing sub-optimal codes is another effective method for reducing false positives.

Attractive directions for follow up research include constructing application-specific codes to increase the throughput of the mFISH experiments, theoretical progress in understanding the optimal assignment of existing codes (e.g. by investigating the geometry of Voronoi sets), extending the generative model and the ensuing channel description to $q$-ary codes (e.g. as in seqFISH and seqFISH+ experimental protocols~\cite{shah2017seqfish,eng2019transcriptome}), and taking into account spatial interference and color channel cross-talk in the data generating process.

\bibliographystyle{IEEEtran}
\bibliography{./project.bib}
\end{document}